\documentclass[%
superscriptaddress,
showkeys,
preprint,
showpacs,preprintnumbers,
 amsmath,amssymb,
aps,
]{revtex4-1}

\usepackage{graphicx}% Include figure files
\usepackage{dcolumn}% Align table columns on decimal point
\usepackage{bm}% bold math
\usepackage{multirow}
\begin{document}

\preprint{APS/123-QED}

\title{Direct reaction measurements with a  $^{132}$Sn radioactive ion beam}% Force line breaks with \\
%\thanks{A footnote to the article title}%

\author{K.L Jones}
\affiliation{%
Department of Physics and Astronomy, University of Tennessee, Knoxville, TN 37996, USA.}
 \affiliation{Department of Physics and Astronomy, Rutgers University, New Brunswick, NJ 08903, USA.}
 
\author{F.M Nunes}
\affiliation{National Superconducting Cyclotron Laboratory and Department of Physics and Astronomy, Michigan State University, East Lansing, MI 48824, USA}

\author{A. S. Adekola}
 \altaffiliation[Current address ]{Department of Physics and Astronomy, Rutgers University, New Brunswick, NJ 08903, USA.}
\affiliation{Department of Physics and Astronomy, Ohio University, Athens, OH 45701, USA.}

 \author{D.W. Bardayan}%
\affiliation{Physics Division, Oak Ridge National Laboratory, Oak Ridge, TN 37831, USA.}%

 \author{J.C. Blackmon}%
 \altaffiliation[Current address ]{Department of Physics and Astronomy, Louisiana State University, Baton Rouge, LA 70803-4001 USA.}%
\affiliation{Physics Division, Oak Ridge National Laboratory, Oak Ridge, TN 37831, USA.}%

  \author{K.Y. Chae}%
   \altaffiliation[Current address ]{Physics Division, Oak Ridge National Laboratory, Oak Ridge, TN 37831, USA.}%
\affiliation{Department of Physics and Astronomy, University of Tennessee, Knoxville, TN 37996, USA.}%

   \author{K.A. Chipps}%
\affiliation{Physics Department, Colorado School of Mines, Golden, CO 80401, USA.}%

\author{J.A. Cizewski}
\affiliation{Department of Physics and Astronomy, Rutgers University, New Brunswick, NJ 08903, USA.}

\author{L. Erikson}
\altaffiliation[Current address ]{Pacific Northwest National Laboratory, P.O. Box 999, Richland, WA 99352, USA}
\affiliation{Physics Department, Colorado School of Mines, Golden, CO 80401, USA.}%

\author{C.~Harlin}
\affiliation{Department of Physics, University of Surrey, Guildford, Surrey, GU2 7XH, UK}

\author{R.~Hatarik}
\altaffiliation[Current address ]{Lawrence Livermore National Laboratory, Livermore,  CA 94550 USA}
\affiliation{Department of Physics and Astronomy, Rutgers University, New Brunswick, NJ 08903, USA.}%

\author{R.~Kapler}
\affiliation{Department of Physics and Astronomy, University of Tennessee, Knoxville, TN 37996, USA.}
\author{R.L.~Kozub}
\affiliation{Department of Physics, Tennessee Technological University, Cookeville, TN 38505, USA}
\author{J.F.~Liang}
\affiliation{Physics Division, Oak Ridge National Laboratory, Oak Ridge, TN 37831, USA.}%
   \author{R.~Livesay}%
     \altaffiliation[Current address ]{Physics Division, Oak Ridge National Laboratory, Oak Ridge, TN 37831, USA.}%
     \affiliation{Physics Department, Colorado School of Mines, Golden, CO 80401, USA.}%

\author{Z.~Ma}
\affiliation{Department of Physics and Astronomy, University of Tennessee, Knoxville, TN 37996, USA.}
\author{B.~Moazen}
%\altaffiliation[Current address ]{Physics Division, Oak Ridge National Laboratory, Oak Ridge, TN 37831, USA.}%
\affiliation{Department of Physics and Astronomy, University of Tennessee, Knoxville, TN 37996, USA.}

\author{C.D~Nesaraja}
\affiliation{Physics Division, Oak Ridge National Laboratory, Oak Ridge, TN 37831, USA.}%

\author{S.D.~Pain}
\altaffiliation[Current address ]{Physics Division, Oak Ridge National Laboratory, Oak Ridge, TN 37831, USA.}%
\affiliation{Department of Physics and Astronomy, Rutgers University, New Brunswick, NJ 08903, USA.}

\author{N.P.~Patterson}
\affiliation{Department of Physics, University of Surrey, Guildford, Surrey, GU2 7XH, UK}
\author{D.~Shapira}
\affiliation{Physics Division, Oak Ridge National Laboratory, Oak Ridge, TN 37831, USA.}%
\author{J.F.~Shriner Jr.}
\affiliation{Department of Physics, Tennessee Technological University, Cookeville, TN 38505, USA}
\author{M.S.~Smith}
\affiliation{Physics Division, Oak Ridge National Laboratory, Oak Ridge, TN 37831, USA.}%
\author{T.P.~Swan}
\affiliation{Department of Physics and Astronomy, Rutgers University, New Brunswick, NJ 08903, USA.}
\affiliation{Department of Physics, University of Surrey, Guildford, Surrey, GU2 7XH, UK}
\author{J.S.~Thomas}
\altaffiliation[Current address ]{Schuster Laboratory, University of Manchester, Manchester, M13 9PL, UK}
\affiliation{Department of Physics, University of Surrey, Guildford, Surrey, GU2 7XH, UK}

\date{\today}

% It is always \today, today,
             %  but any date may be explicitly specified

\begin{abstract}
The (d,p) neutron transfer and (d,d) elastic scattering reactions were measured in inverse kinematics using a radioactive ion beam of $^{132}$Sn at 630 MeV.  The elastic scattering data were taken in a region where Rutherford scattering dominated the reaction, and nuclear effects account for less than 8$\%$ of the cross section.  The magnitude of the nuclear effects was found to be independent of the optical potential used, allowing the transfer data to be normalized in a reliable manner.  The neutron-transfer reaction populated a previously unmeasured state at 1363~keV, which is most likely the single-particle 3p$_{1/2}$ state expected above the $N=82$ shell closure.  The data were analyzed using finite range adiabatic wave calculations and the results compared with the previous analysis using the distorted wave Born approximation.  Angular distributions for the ground and first excited states are consistent with the previous tentative spin and parity assignments.  Spectroscopic factors extracted from the differential cross sections are similar to those found for the one neutron states beyond the benchmark doubly-magic nucleus $^{208}$Pb.
%\begin{description}
%\item[Usage]
%Secondary publications and information retrieval purposes.
%\item[PACS numbers]
%\pacs{#1} 25.60.Je (transfer reactions with unstable nuclei) 25.60.Bx (Elastic Scattering with unstable nuclei) 25.45.Hi (transfer reactions, deuteron) 29.38.Gj (reaccelerated RIBs)
%\end{description}
\end{abstract}

% PACS, the Physics and Astronomy
                             % Classification Scheme.
%Use showkeys class option if keyword
\maketitle
\section{\label{sec:intro}Introduction\protect\\ }
A subject of great interest in nuclear structure physics is the evolution of single-particle structure far from stability and how this can reflect changes in shell structure.  Transfer reactions are a powerful probe for investigating the single-particle structure of nuclei \cite{Gle75,Sat83}.  Using solid targets with a high stoichiometry of protons or deuterons, for example, it is possible to perform transfer reactions in inverse kinematics on beams of nuclei which cannot easily be made into targets  \cite{Kra91}.  With the availability of beams of short-lived ions at energies relevant to these studies, i.e. of a few MeV to a few tens of MeV per nucleon,  it is now possible to conduct transfer reactions with short-lived nuclei in inverse kinematics (for example \cite{Win97, Reh98, tho05, tho07,Wim10}).  In particular, the single-neutron stripping reaction (d,p) can be performed using beams of short-lived fission fragments impinging on targets of deuterated plastic \cite{tho05,tho07}.  The experiments described here represent the first direct reaction measurements using a  beam of short-lived, t$_{1/2}$ = 40 s, $^{132}$Sn at energies close to the Coulomb barrier.

As transfer reactions selectively populate single-particle or single-hole states, they have particular relevance close to shell closures.  With $Z = 50$ and $N = 82$, $^{132}$Sn belongs to the select group of nuclei with standard magic numbers of both protons and neutrons.  This is seen, for example, through the high energy of the first 2$^+$ state in the tin isotopes (around 1.2~MeV), compared to the neighboring elements (typically around 500~keV).  Additionally, the large discontinuities seen in neutron-rich tin isotopes in both the first 2$^+$ energy, rising to 4.04~MeV in $^{132}$Sn, and  the two neutron separation energy, S$_{2n}$, falling from 12.56~MeV for $^{132}$Sn to 6.38~MeV for $^{134}$Sn (N~=~84),  are indicative of the doubly-magic nature of $^{132}$Sn.

\begin{figure}
\includegraphics[width=7.5cm]{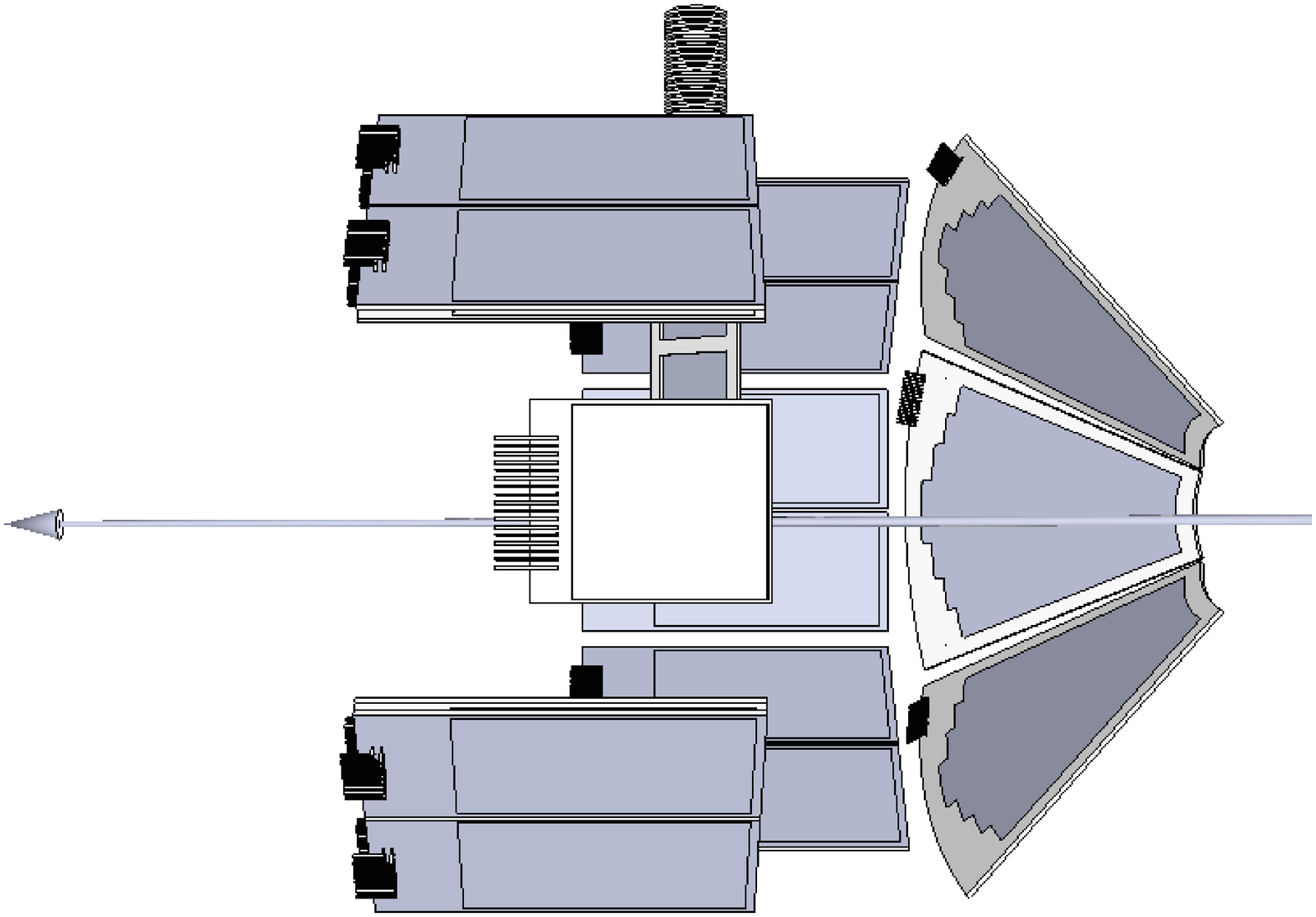}
\includegraphics[width=5.5cm]{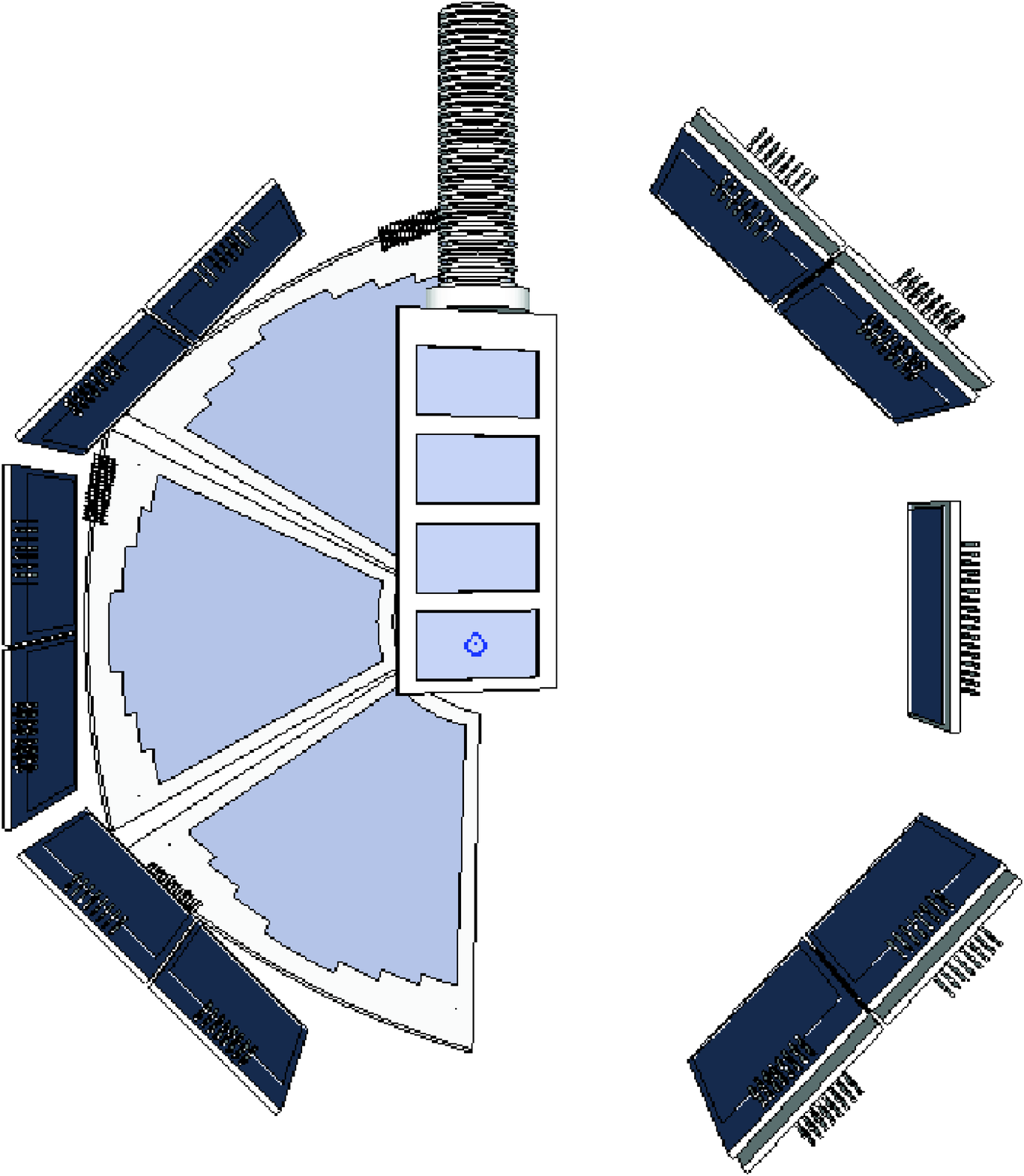}
\caption{\label{setupfig}Diagrams from the side (top panel) and beam views (bottom panel) of the experimental setup (not to scale) \protect\cite{Sch11}.  }
\label{setup}
\end{figure}

The characterization of the states in $^{133}$Sn is critical to understanding the evolution of single-particle structure in this important region of the nuclear chart and to extrapolating the properties of nuclei outside the current reach of detailed experimental study, including those on the astrophysical r-process path.  The (d,p) reaction at energies around the Coulomb barrier favorably populates low-energy, low-angular momentum single-particle states.  The lowest-energy neutron-particle states expected above the $N=82$ shell closure are 2f$_{7/2}$, 3p$_{3/2}$, 3p$_{1/2}$, 1h$_{9/2}$, 2f$_{5/2}$, and 1i$_{13/2}$.   Previous to the current experiment, candidates for all but the 3p$_{1/2}$ and the 1i$_{13/2}$ states had been identified in $^{133}$Sn \cite{Hof96,Urb99}.

Previous studies of $^{133}$Sn have used $\beta$ decay \cite{Hof96} or the prompt $\gamma$-decay of fission fragments from $^{248}$Cm \cite{Urb99}.  The $\beta$ decay of $^{133}$In is dominated by the ${(\pi g_{9/2})}^{-1}$ configuration, which favors the population of states with spin $7/2$, $9/2$ or $11/2$.  Similarly, low spin states are disfavored in the $\beta$-delayed neutron decay of $^{134}$In; typically, states within one unit of angular momentum to those observed from the $\beta$ decay are populated.  Since the fission process favors fragments with significant angular momentum, the prompt gamma rays observed are associated with decay of higher spin excitations on or near the yrast line.  Hence, previous studies of $^{133}$Sn most likely missed low-spin states. 

%The methods of identifying $\gamma$ rays following fission involve gating on high multiplicity events usually associated with high spin states. Although most of the bound states in $^{133}$Sn were identified via these methods, the $1/2^-$ state had not been observed.  A $(1/2)^-$ state at 1656~keV, presumably the single particle p$_{1/2}$ state, was commonly referred to in the literature despite the fact that it was only very tentatively suggested as a possible state in the original paper \cite{Hof96}.

The information that can be gained from transfer-reaction experiments includes energies and angular distributions of the light-ion ejectiles. The energies provide the excitation energies of the heavy recoil, and the angular distributions can be used to extract the $\ell$ value of the transfered nucleon.  By comparison of the differential cross sections for individual states with those calculated using a reaction model, spectroscopic factors can be extracted.  In the region which is well described by Rutherford scattering, elastically scattered target components can be used to accurately normalize the data from transfer.  Additionally, the elastic scattering of beams of exotic nuclei could be used to improve optical model potentials away from stability.  Some of the results of the $^{132}$Sn(d,p) reaction study have been previously reported \cite{Jon10}.  The present paper provides a more detailed presentation of the experimental results, including elastic scattering of deuterons, as well as an interpretation of the data within the ADiabatic Wave Approximation (ADWA).

\section{\label{sec:level1}Experimental Setup\protect\\ }
The experiment was performed at the Holifield Radioactive Ion Beam Facility \cite{Bee11} at Oak Ridge National Laboratory.  The $^{132}$Sn ions were produced from the fission of $^{nat}$U following the bombardment of protons (up to a maximum energy of 50 MeV) using the isotope separation online (ISOL) technique.  The beam was purified by extracting tin sulfide molecules and selecting mass = 164 at the first stage separator.  Following charge exchange and subsequent breakup of the SnS molecules, $^{132}$Sn ions were accelerated to a total energy of 630~MeV in the 25~MV tandem accelerator.  The essentially pure ($>$ 90$\%$) $^{132}$Sn beam impinged on a deuterated polyethelene target  with areal density of 80~$\mu$g/cm$^2$.  The target was turned 30$^{\circ}$ to the beam axis, resulting in an effective target thickness of 160~$\mu$g/cm$^2$,  to allow the measurement of emitted particles close to $\theta_{lab} = 90^{\circ}$.   

Scattered light ions and protons emitted from the (d,p) reaction were measured in a system of silicon detectors incorporating the Silicon Detector Array (SIDAR \cite{Bar99}) and an early implementation of the Oak Ridge Rutgers University Barrel Array (ORRUBA) \cite{Pai07}, as shown in Fig. \ref{setup}.  SIDAR, placed at backward angles in the laboratory frame for a subsequent experiment, was exposed to a small number of reaction protons as only the population of $\ell = 0$ states would emit significant numbers of protons at these backward laboratory angles owing to the inverse kinematics of the experiment.  The ORRUBA detectors covered polar angles between 69$^{\circ}$ and 107$^{\circ}$.  At angles forward of $\theta_{lab}$ = 90$^{\circ}$, where the detectors were exposed to elastically scattered protons, deuterons and $^{12}$C target constituents, four telescopes of ORRUBA detectors were employed.  Three of the telescopes used 140~$\mu$m  $\Delta$E (energy loss) detectors, and the other used a 65~$\mu$m  $\Delta$E detector.  The second layer of the telescopes, and single layer detectors backward of the elastic scattering region, were 1000~$\mu$m thick. The high capacitance of both types of $\Delta$E detectors, combined with the necessary resistive layer for charge division, resulted in low signal-to-noise ratios and incomplete charge collection.  Subsequent to this measurement non-resistive 65~$\mu$m  $\Delta$E detectors have been incorporated into the ORRUBA setup.  Owing to the poor resolution of the $\Delta$E detectors, the signals were not used in the analysis of the transfer data.  Instead the energy loss in these detectors was reconstructed from the measured residual energy in the E detectors.  This impacted the resulting Q-value resolution, leading to a resolution of $\approx~$300 keV.
\begin{figure}
\includegraphics[width=9.6cm]{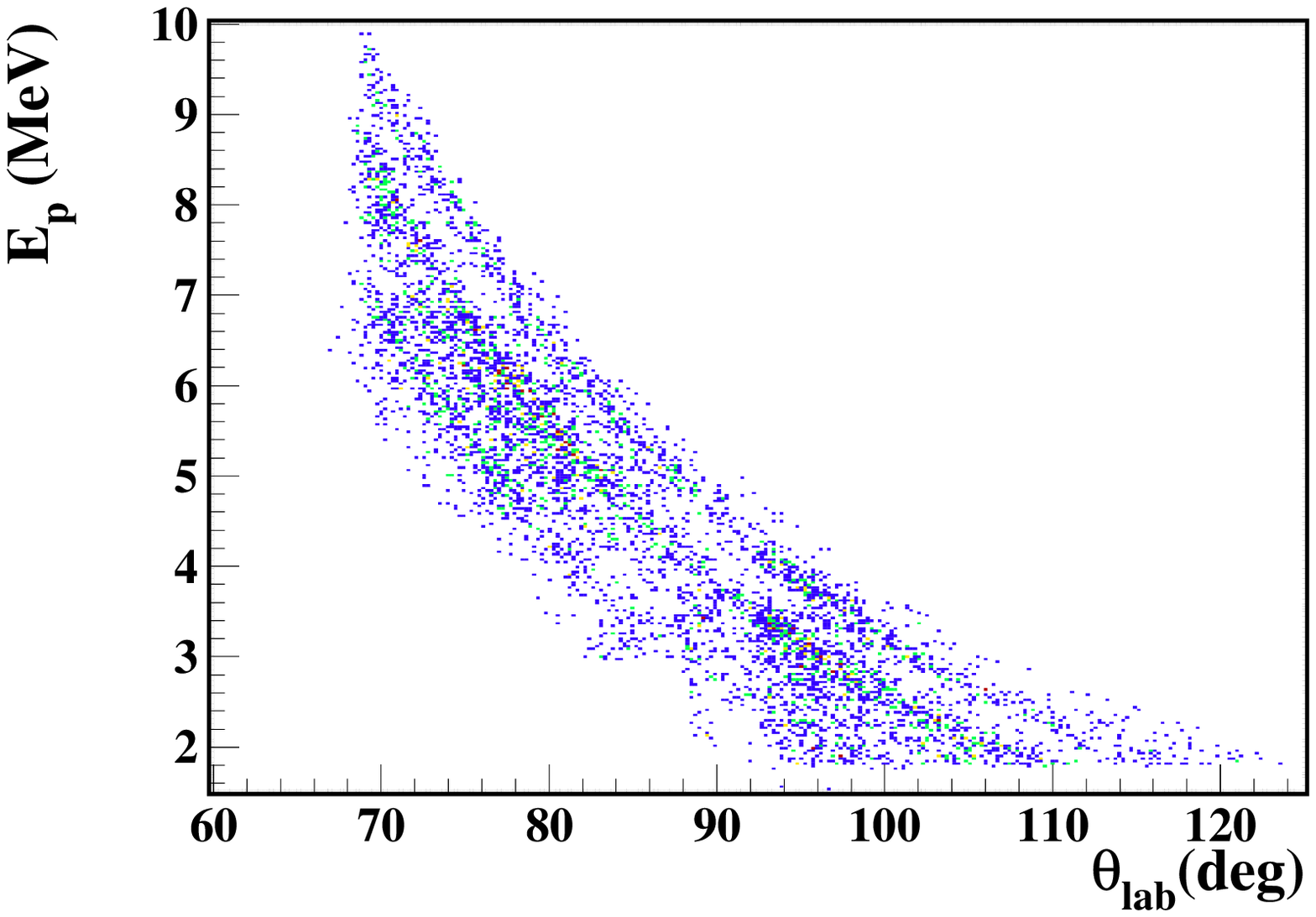}
\includegraphics[width=9.6cm]{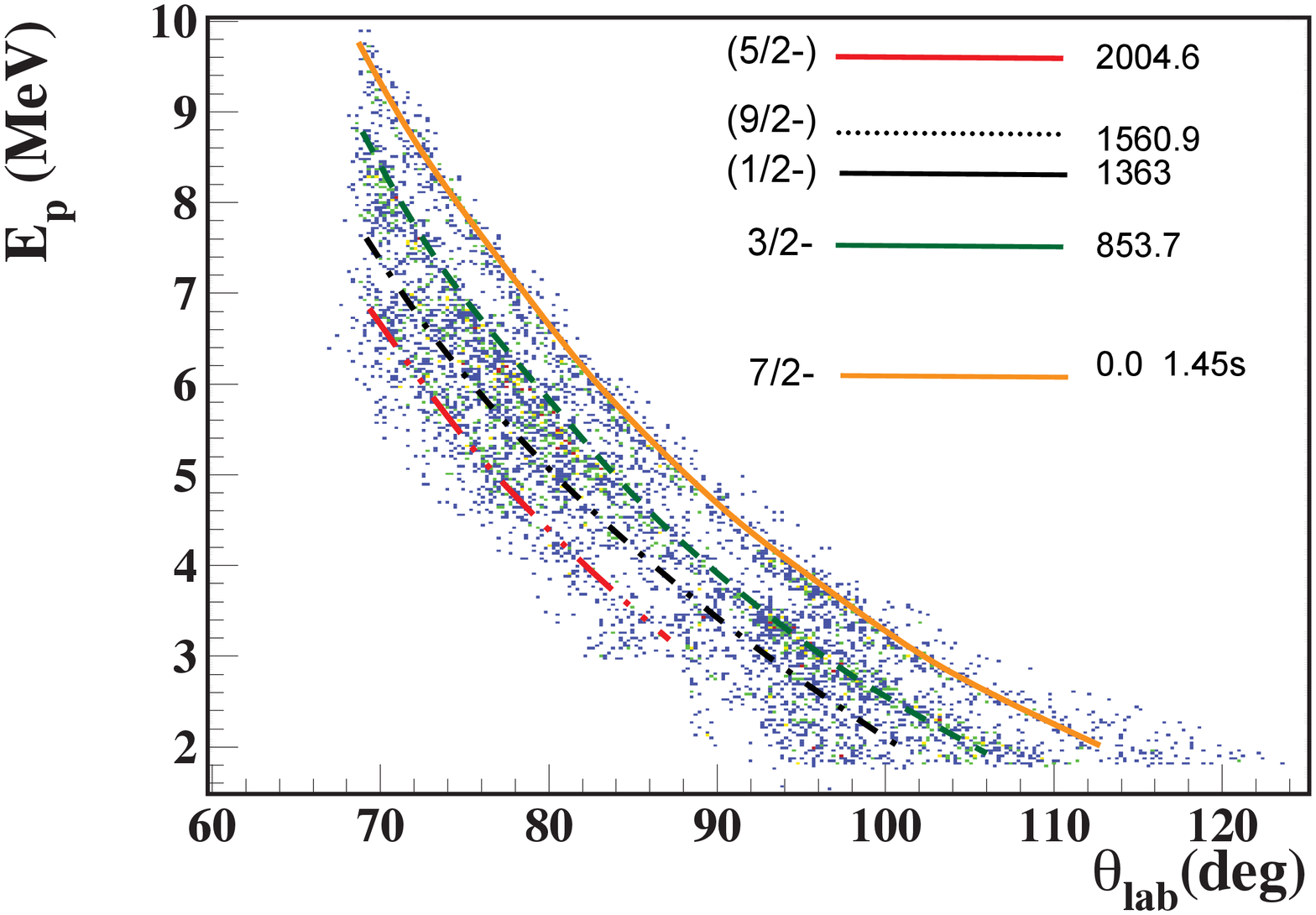}
\caption{\label{ethetafig}Energy versus angle measurements for the protons emitted from the $^{132}$Sn(d,p) measurement.  Equi-Q-value lines are shown to guide the eye in the lower panel for the ground (yellow, solid), 854-keV (green, dashed), 1363-keV (black, dot-dashed) and 2005-keV (red, dot-dot-dashed) states.  The inset shows the low-energy states in $^{133}$Sn (color online).}
\label{etheta}
\end{figure}
The energies and angles of protons emitted from the (d,p) reaction follow well defined loci dependent on the Q-value of the reaction, as shown in Fig.~\ref{etheta}.   Lines representing the calculated kinematic loci for reactions resulting in $^{133}$Sn being produced in its ground, 854-, 1363-, and 2005-keV states are shown to help guide the eye.
\begin{figure}
\includegraphics[width=10cm]{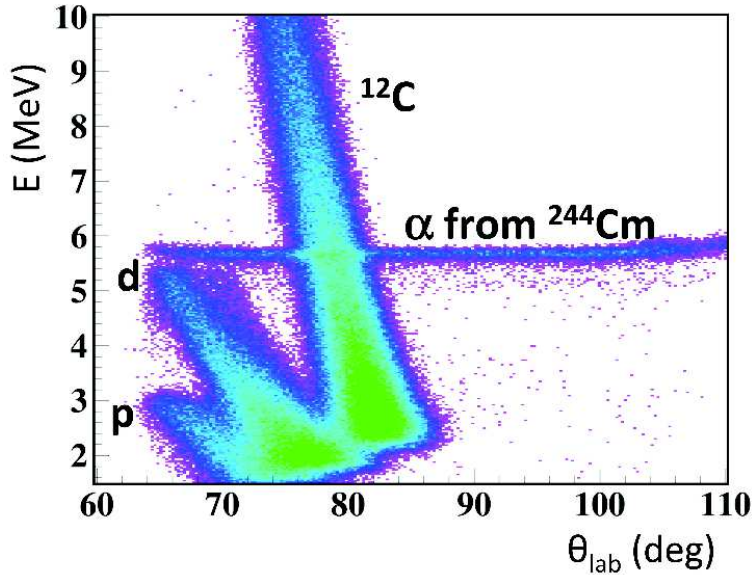}
\caption{\label{elasticfig}Energy versus angle measurements for protons, deuterons and $^{12}$C atoms scattered out of the deuterated polyethylene target, measured in a single $\Delta$E detector.  The horizontal line is from the $\alpha$ decay of $^{244}$Cm present in the target chamber (color online).}
\label{elastic-2D}
\end{figure}
\section{\label{sec:elastics}Elastic scattering of $^{132}$S\lowercase{n} on a deuteron target\protect\\}
The purpose of measuring the elastic scattering of the $^{132}$Sn beam on a deuteron target was two-fold. Most importantly, these data can provide a robust method of normalization of the transfer cross sections.  Secondly, in general, elastic scattering data can also be used to constrain the parameters of the optical potential.  In order to normalize the transfer cross sections, it was important to obtain data in the region where Rutherford scattering dominates, i.e. close to 0$^{\circ}$ in the center-of-mass system, corresponding to 90$^{\circ}$ in the laboratory system.  Ideally, to fit the optical potential, a larger range of center-of-mass angles should be covered in the elastic scattering data.

\begin{figure}
\includegraphics[width=6.0cm, angle=270]{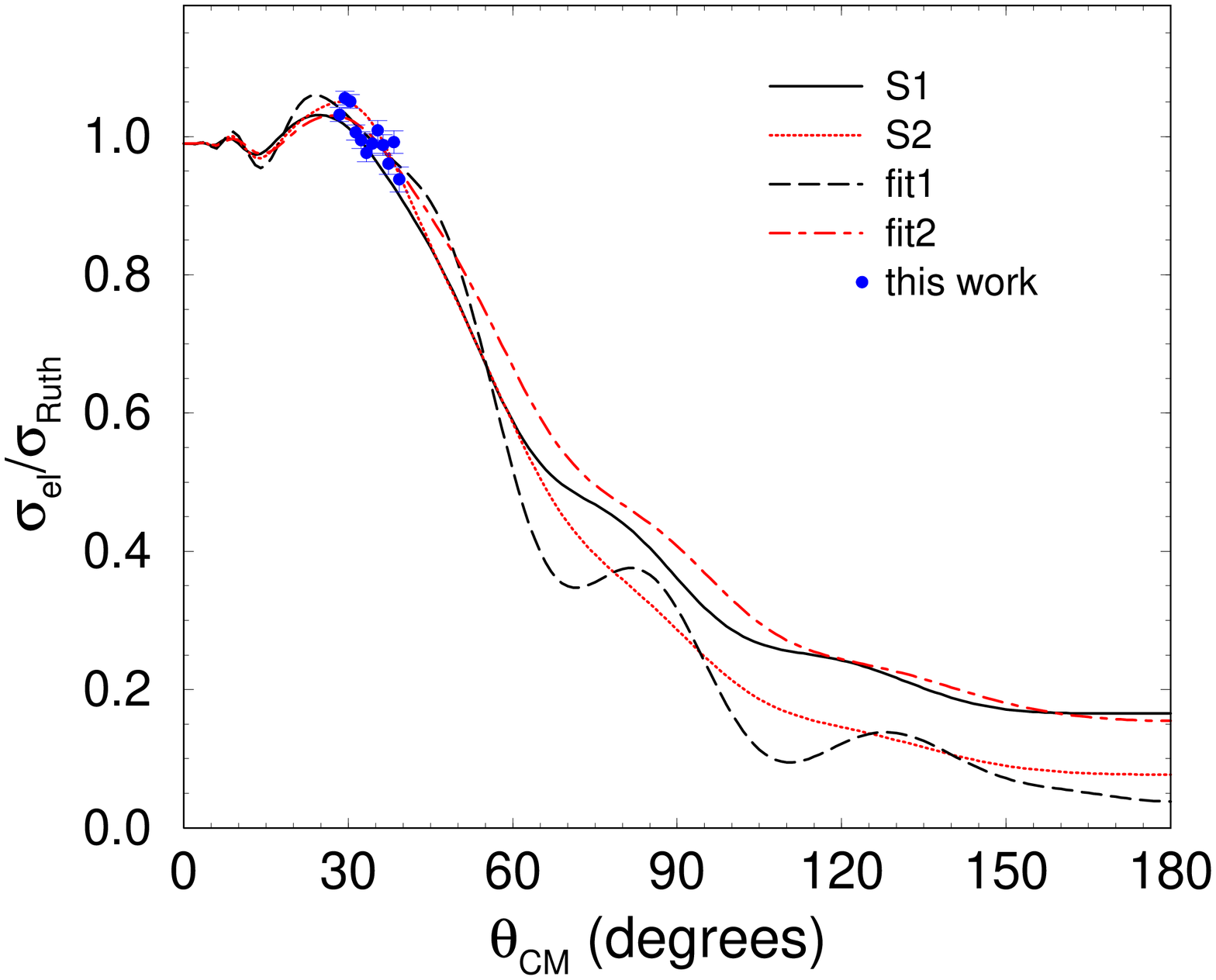}
\includegraphics[width=6.0cm, angle=270]{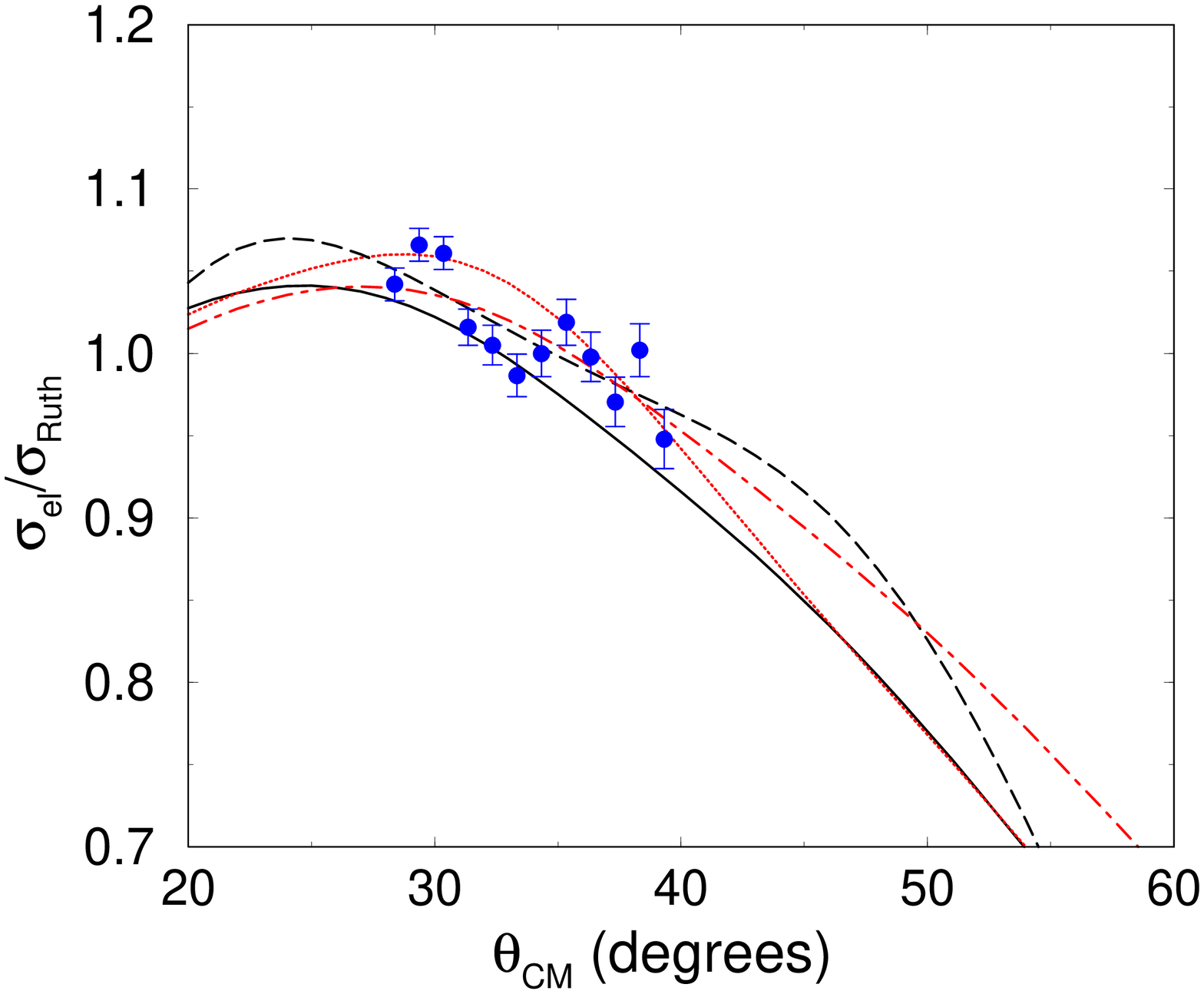}
\caption{\label{elastic-fit}Ratio of measured to Rutherford cross section s for the $^{132}$Sn(d,d) reaction. The solid curves are calculations using potentials from Str{\"o}mich et al. \cite{Str77}, one with a surface imaginary term (S1 black, solid) and the other with a volume imaginary term (S2 red, dotted). Examples of the result of $\chi^2$ fitting to S1 and S2 are shown as fit 1(black, dashed) and fit 2 (red, dot-dashed)respectively.  The right-hand panel shows an expanded view of the region covered by the data (color online).}
\label{elastics}
\end{figure}

The elastically scattered deuterons were measured in the 140~$\mu$m thick $\Delta$E detector with the best performance.  The most forward center-of-mass angles correspond to the lowest energy elastically scattered particles.  Therefore, the angle and energy resolutions degrade appreciably at center-of-mass angles much below $\theta_{CM}=30^{\circ}$ ($\theta_{lab}=75^{\circ}$), as shown in Fig.~\ref{elastic-2D}.  Data from the $^{132}$Sn(d,d) reaction were extracted for 28.4$^{\circ}\le\theta_{CM}\le 39.3^{\circ}  ($ 70.4$^{\circ}\le\theta_{lab}\le 75.8^{\circ}$) (Fig.~\ref{elastics}).  Over this range of angles, the excursions from pure Rutherford scattering were less than 8\%.  In order to be able to use the elastic scattering data to normalize the transfer cross sections, the contribution coming from nuclear scattering had to be taken into account.

In Fig.\ref{elastics}, in comparison with the data, the calculated angular distributions using two optical potentials based on the $^{124}$Sn deuteron optical potential from Str\"{o}mich et al. \cite{Str77} are displayed.  Set S1 includes a surface imaginary term and S2 includes a volume imaginary term. Considering a 5\% uncertainty in the normalization of the data, fits for both cases result in a $\chi^2 \approx 1$. While both S1 and S2 may seem like adequate choices for the range where data are available, they differ significantly over a wider angular range, and it is not clear that either would be adequate to describe elastic scattering of deuterons off $^{132}$Sn (note that the potentials from \cite{Str77} were obtained from data on stable isotopes including backward angles). Subsequently, a series of fits of the optical potential parameters were performed.  Shown in Fig.\ref{elastics} are the results of two such fits (fit1 based on S1 and fit2 based on S2). 
Not surprisingly, the fits demonstrated that the optical potential parameters are not well constrained by this narrow angular range. Nevertheless, the resulting elastic scattering in this range does not change by more than 5\% due to the dominance of the Coulomb potential. Set S2 was used to normalize the (d,p) data. A systematic uncertainty of 5\% coming from  this normalization has been applied to the cross sections from the transfer data.

\section{\label{sec:dp}The  $^{132}$S\lowercase{n}(\lowercase{d,p}) reaction in inverse kinematics\protect\\ }
\begin{figure}
\includegraphics[width=9.0cm]{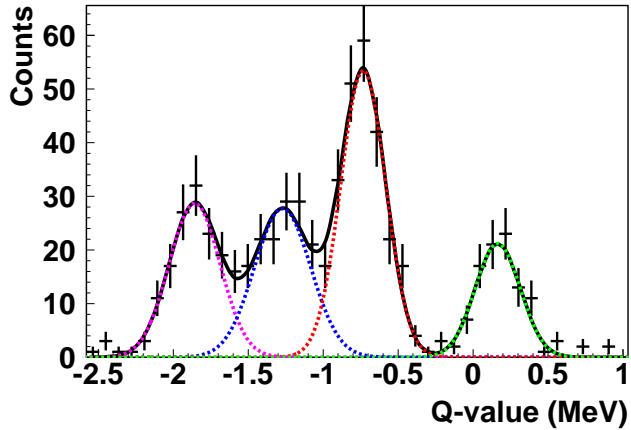}
\caption{\label{qfig}Q-value spectrum for the $^{132}$Sn(d,p)$^{133}$Sn reaction with a 630~MeV $^{132}$Sn beam.  The solid black line shows a fit including four peaks: the ground state (green), the 854 keV state (red), the previously unobserved 1363 keV state (blue) and the 2005 keV state (magenta) (color online).}
\label{q}
\end{figure}

\begin{figure*}[ht]
\includegraphics[width=8.0cm, angle=270]{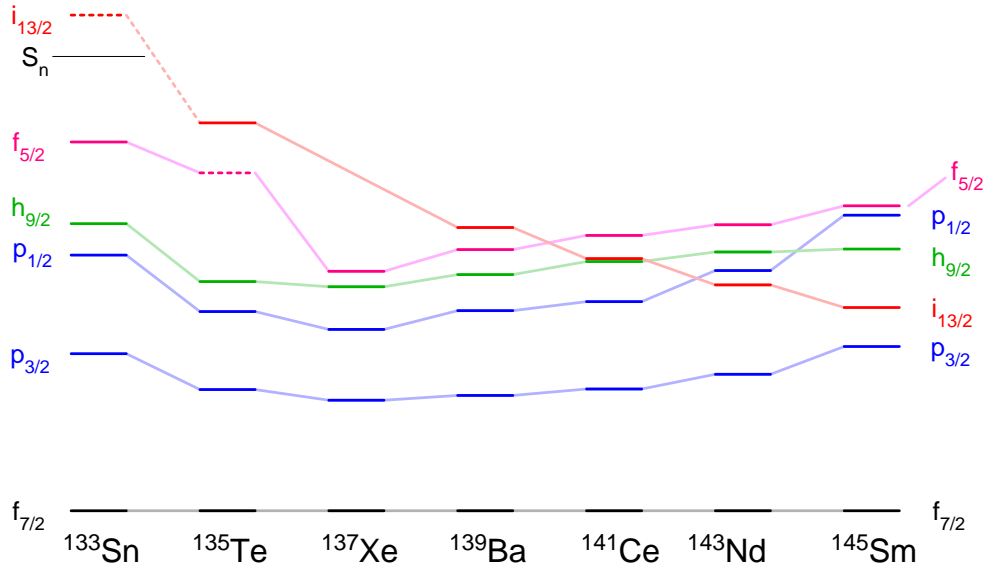}
\caption{\label{n83}Single-particle states in the $N=83$ even-Z isotones between $^{145}$Sm and $^{133}$Sn.  Data taken from the Evaluated Nuclear Structure Data File \cite{ENSDF}, \cite{Pai11}, and present results (color online).}
\label{n83fig}
\end{figure*}

The Q-value for population of $^{133}$Sn via the (d,p) reaction was calculated on an event-by-event basis from the measured angle and energy of the emitted proton.  The energy of the beam was corrected for energy loss in the target, assuming that the reaction occurred in the center of the target, as was the energy of the proton.  Three states had been previously observed in $^{133}$Sn:  the ground state, 854- and 2005-keV excited states with tentative spin assignments of 7/2$^-$, 3/2$^-$ and 5/2$^-$, respectively. Four peaks were observed in the Q-value spectrum (Fig. \ref{q}) corresponding to these states, as well as a newly observed state.  The excitation energy of this new state, interpolated from the Q-values of the other three peaks, was determined to be $1363\pm31$~keV.  Figure \ref{n83fig} summarizes the systematics of single-particle states in N~=~83 isotones, including the new 1363-keV candidate for the p$_{1/2}$ state. The smooth variation of the energy of 1/2$^-$ states in $N~=~83$ isotones, including the candidate in $^{133}$Sn, supports this assignment; however, more evidence is required before the assignment can be considered firm.

%Four peaks were observed in the Q-value spectrum (Fig.~\ref{q}) corresponding to the ground state, the first excited state at E$_x$~=~854~keV, a previously unobserved state and a previously measured state at E$_x$~=~2005~keV.  The excitation energy of the newly observed state, interpolated from the known Q-values of the other three peaks, was found to be 1363~$\pm$31keV.  Fig.~\ref{n83} shows this energy level plotted with the single-particle states previously measured in even-N neutron-rich tin isotopes assuming that this state is indeed the previously unobserved p$_{1/2}$ single-particle state.  
\begin{figure}
\includegraphics[width=6.0cm]{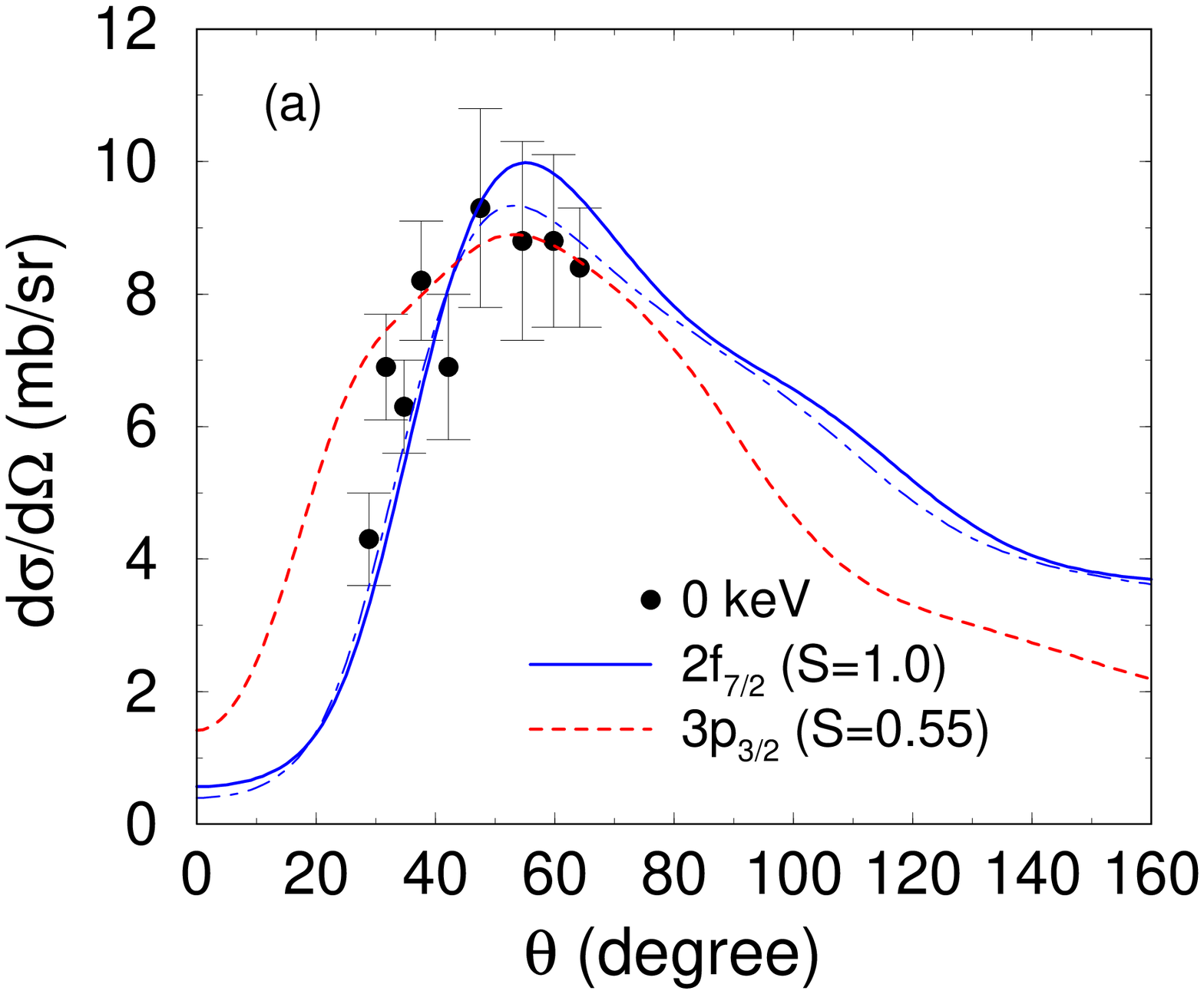}
\includegraphics[width=6.0cm]{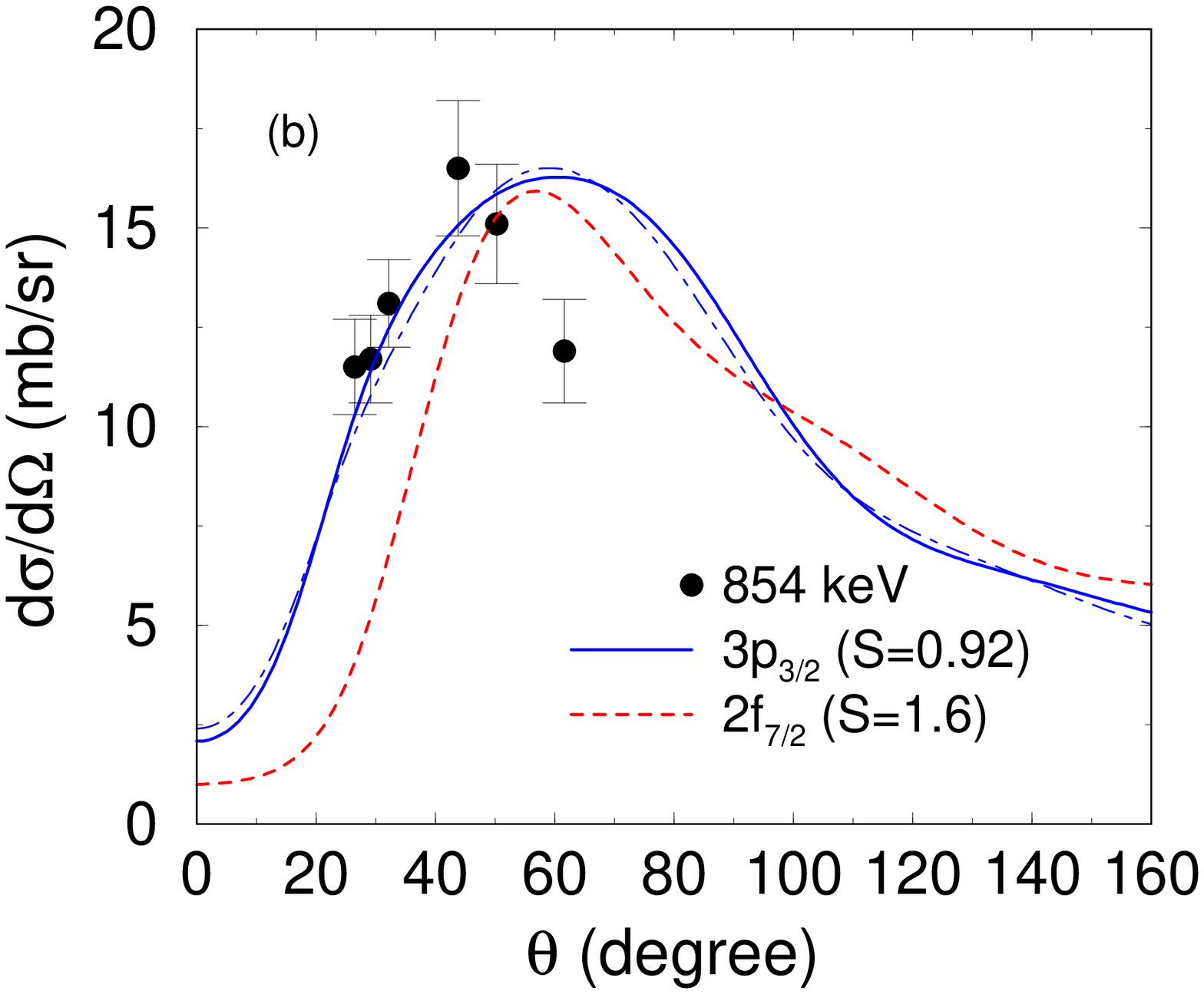}
\includegraphics[width=6.0cm]{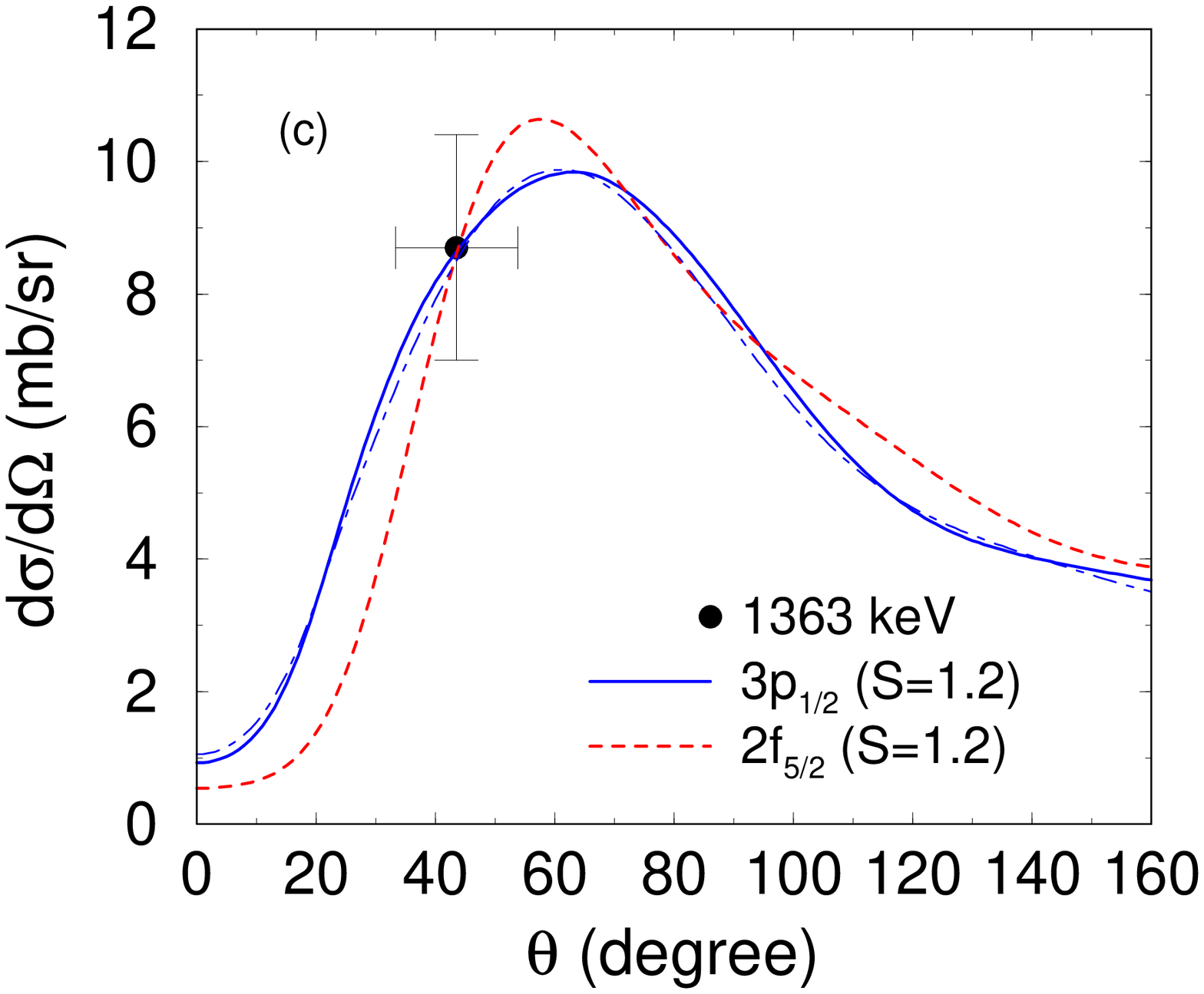}
\includegraphics[width=6.0cm]{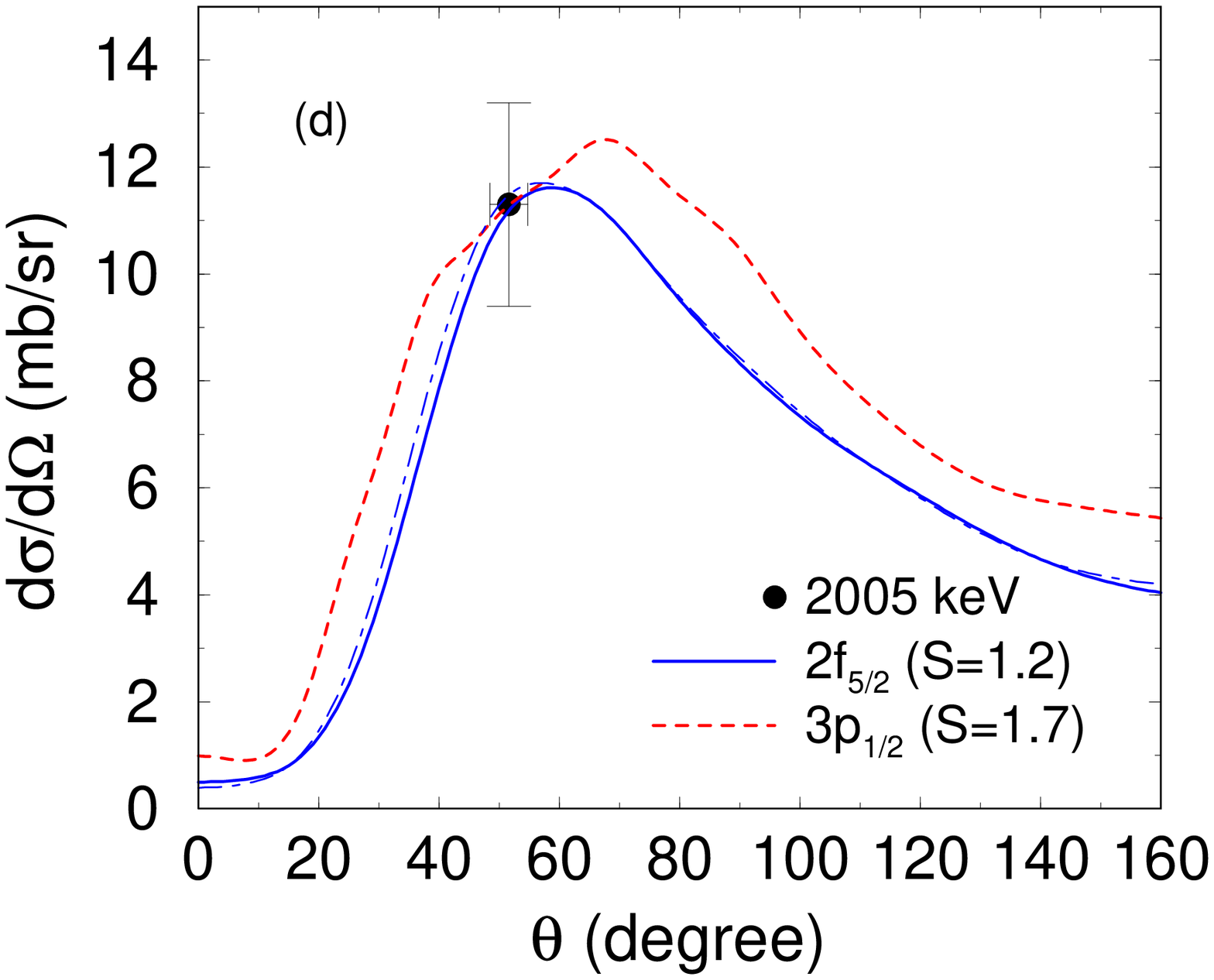}
\caption{\label{ang}Angular distributions of protons in the center of mass from the $^{132}$Sn(d,p)$^{133}$Sn reaction for the two lowest states populated and integrated cross section  measurements for the two highest states.  Calculations assuming the n$\ell$j assignments in Table \ref{tab1}, using ADWA-CH, are shown in blue (solid).  Similar calculations assuming the nearest expected alternate $f$-wave or $p$-wave single-neutron state are shown in red (dotted).  The numbers in parentheses give the spectroscopic factors used to fit the calculation to the data. Calculations using ADWA-BG assuming the n$\ell$j assignments in Table \ref{tab1} are shown in blue (dot-dashed):   a. Ground state, b. 854-keV state, c. 1363-keV state, and d. 2005-keV state.(color online).}
\label{angdist}
\end{figure}

To confirm single-particle assignments for the ground and first excited states, angular distributions of protons emitted following the (d,p) reaction were measured (see Fig.~\ref{angdist}) and compared to calculations assuming either an $\ell = 1$ or $\ell = 3$ transfer (these calculations are described in Section \ref{sec:analysis}).  It should be noted only $p$-wave or $f$-wave states are expected to be significantly populated using a low-energy (d,p) reaction in N~=~83 nuclei near $^{132}$Sn.

The angular distributions were extracted from the data by fitting four gaussians to the four states shown in Fig.~\ref{q} at each angle.  Figure~\ref{angdist}(a) shows the measured angular distribution of protons following population of the ground state, in the center-of-mass, and calculations assuming transfer to the 2f$_{7/2}$ and 3p$_{3/2}$ states.  Both calculations agree with the ground-state data across most of the range of angles; however, the $\ell = 3$ calculation is preferred at the most forward angles (around $\theta_{CM}=20^{\circ}$).  For the 854-keV state, (Fig. \ref{angdist}(b)) an $\ell=1$ transfer reproduces the data, particularly at more forward angles.  The assignments of the 2f$_{7/2}$ and 3p$_{3/2}$ states are therefore confirmed.

Population of the higher excited levels in $^{133}$Sn led to lower energy protons being detected in the ORRUBA detectors.  As these detectors use charge division to extract position information, it was necessary to receive a significant signal from either end of the detector in order to measure the energy and position of the charged particle.  This became increasingly difficult at lower proton energies.  Additionally, if the proton strikes near one end of the strip, the signal at the far end is degraded in quality. In particular it is spread out in time, which can lead to incomplete charge collection or, in the worst cases, the signal not arriving within the time required by the acquisition gate.  For these reasons, only the central portion of the ORRUBA detectors was able to provide data for the low-energy protons emitted from the population of the 1363- and 2005-keV states. Hence angle-integrated cross sections, rather than angular distributions, are shown in Figs. \ref{angdist}(c) and (d).  The 2005-keV state had been previously observed in beta-decay and assigned (5/2$^-$), consistent with population from a high-spin parent. If indeed this is the expected 2f$_{5/2}$ level, the spectroscopic factor is consistent with 1.  An alternative 3p$_{1/2}$ assignment is inconsistent with its observation in beta decay.  On the other hand, the new 1363-keV state had not been previously observed via beta-decay, beta-delayed neutron decay \cite{Hof96} or in the gamma decay following fission \cite{Urb99}.  Therefore it is likely that this is a lower-spin state and a good candidate for the expected 3p$_{1/2}$ state.  The assignments of the 1363-keV state as 3p$_{1/2}$ and 2005-keV state as 2f$_{5/2}$ are also supported by the systematics summarized in Fig. \ref{n83fig}.     %Therefore, it was not possible to provide experimental evidence in this measurement that the 1353~keV state is indeed the missing p$_{1/2}$ state except through the elimination of other possibilities.  However, this state was not previously observed via $\beta$-decay, $\beta$-delayed neutron-decay \cite{Hof96} or in the $\gamma$ rays following fission \cite{Urb99}, each of which is expected to be sensitive to higher angular momentum states.  This strongly suggests that the 2005~keV state observed in those experiments is the 2f$_{5/2}$ and the 1363~keV state is the 3p$_{1/2}$, supported by the systematics shown in Fig~\ref{n83}.

\section{\label{sec:analysis}Reaction analysis\protect\\ }
The initial analysis \cite{Jon10} of the $^{132}$Sn(d,p) reaction was performed in the finite-range distorted-wave Born approximation (DWBA) framework, with the code FRESCO \cite{tho88}.  The optical model parameters were taken from Str\"{o}mich et al. \cite{Str77}. For completeness these results for the spectroscopic factors and asymptotic normalization coefficients are summarized in Tables \ref{tab1} and \ref{tab2}, respectively.  The same framework was also used to analyze the $^{208}$Pb(d,p) reaction.  Those results, and results from choosing global optical model parameters were also reported in \cite{Jon10} and the associated supplemental material. 

Although traditionally the one-step DWBA has been widely used to analyze (d,p) reactions, it has been long known that it is often inaccurate. Johnson and Soper \cite{Joh70} showed the importance of including deuteron breakup explicitly and devised a practical method for analyzing (d,p) reactions which is non-perturbative, the so-called ADiabatic Wave Approximation. Its simplicity arises partly from the use of the zero-range approximation for the deuteron. A finite-range version of this method \cite{Joh74} has recently been applied to a wide range of (d,p) reactions \cite{Ngu10}, showing the importance of performing a full finite-range calculation. This method, referred to as FR-ADWA, is the method used here to analyze all of
the transfer data. Within this method, the deuteron adiabatic wave is constructed from nucleon optical potentials, reducing considerably the optical parameter uncertainties. 

\begin {table}[t]
\caption{\label{tab1}
Spectroscopic factors, S, of the four single-particle states populated by the $^{132}$Sn(d,p)$^{132}$Sn reaction extracted using DWBA \cite{Jon10} and ADWA formalisms.  Quoted error margins include only experimental uncertainties.  The values extracted from the ADWA-CH are considered the most reliable and are shown in bold.}

\begin{ruledtabular}
\begin{tabular}{ccccc}
& & \multicolumn{3}{c}{Spectroscopic Factor} \\
\cline{3-5}
\textrm{E$ _x (keV)$}&
\textrm {n$\ell$j} &
\textrm{DWBA} &
\textrm{FR-ADWA-BG} &
\textrm{\bf{FR-ADWA-CH}} \\

\colrule
0 			& 2f7/2	& $0.86 \pm0.07$ 	& $1.2 \pm 0.1$ 	& \mbox{\boldmath{$1.00 \pm0.08$}} 	\\

854 			& 3p3/2 	& $0.92 \pm0.07$ 	& $1.0 \pm 0.1$ 	& \mbox{\boldmath{$0.92 \pm0.07$}} 	\\

1363$\pm31$ 	& (3p1/2) 	& $1.1 \pm0.2$ 	& $1.2 \pm 0.3$ 	& \mbox{\boldmath{$1.2 \pm0.2$}} 	\\

2005 		& (2f5/2)	& $1.1 \pm0.2$ 	& $1.3 \pm 0.3$ 	& \mbox{\boldmath{$1.2 \pm0.3$	}}\\
\end{tabular}
\end{ruledtabular}
\end{table}
Calculations for the $^{132}$Sn(d,p)$^{133}$Sn reaction were performed in post-form using the reaction code {\sc fresco} \cite{tho88}. A realistic Reid interaction \cite{Rei68} 
was used for the deuteron with the neutron-proton potential, $V_{np}$, in the transfer operator. The global optical model parameters CH89 \cite{Var91} were used for all nucleon optical potentials.  The deuteron adiabatic potential was generated with {\sc twofnr} \cite{twofnr}.  One important source of ambiguity in modeling transfer reactions is the choice of the single-particle parameters used for the overlap function, in this case $^{133}$Sn relative to $^{132}$Sn. Even if radii predicted by density functional theory \cite {Ter05} for the tin isotopes were used, the geometry of the valence neutron would still carry large uncertainties. Therefore, the mean field was fixed to a standard Woods-Saxon shape with radius $r=1.25$~fm and diffuseness $a=0.65$~fm.

FR-ADWA angular distributions for single-neutron transfer to all measured states are presented in Fig.~\ref{ang} using the preferred $nlj$ assignment (blue solid line), as indicated in Tables \ref{tab1} and \ref{tab2}, and an alternate $nlj$ assignment (red dashed line).  These assignments represent the nearest expected $\ell=1$ and $\ell=3$ single-particle states. The calculations have been scaled to the data, with the scaling representing the spectroscopic factor, shown in the legend and also summarized in Table~\ref{tab1}. 

The normalization of the many-body overlap function of $^{133}$Sn relative to $^{132}$Sn for large neutron-$^{132}$Sn distances is characterized by the asymptotic normalization coefficient (ANC), usually
denoted by $C$.  The squares of the ANCs, $C^2$, for the various states are summarized in Table~\ref{tab2}.
%obtained from the single particle ANC $b$ by $C^2=S b^2$ \cite{Pan07}.
%Experimental sources of uncertainty in the spectroscopic factors presented in Table~\ref{tab1} and in the ANCs presented in Table~\ref{tab2} are from statistics (4-7\%), fitting (6-20\%) and normalization of the data (5\%). 
\begin {table}[b]
\caption{\label{tab2}
Asymptotic Normalization Coefficients (ANC) of the four single-particle states populated by the $^{132}$Sn(d,p)$^{132}$Sn reaction.  Quoted error margins include only experimental uncertainties. 
}
\begin{ruledtabular}
\begin{tabular}{ccccc}
& &  \multicolumn{2}{c}{C$^2$(fm$^{-1}$)}\\
\cline{3-4}
\textrm{E$ _x (keV)$}&
\textrm {n$\ell$j} &
\textrm{DWBA} &
\textrm{FR-ADWA-CH} \\
%\textrm{C$^2$(fm$^{-1})$} \\

\colrule
0 		& 2f7/2		& $0.64 \pm 0.05$  			&	$0.82 \pm 0.07$\\

854 		& 3p3/2 	 	& $5.6 \pm 0.4$   			&	$6.5 \pm 0.5$\\

1363$\pm31$ 	& (3p1/2) 	& $2.6 \pm 0.6$  			&	$2.9 \pm 0.6$\\

2005 	& (2f5/2)		& $(0.9 \pm 0.2) \times 10 ^{-3}$  		& $(1.2 \pm 0.3)\times 10 ^{-3}$\\
\end{tabular}
\end{ruledtabular}
\end{table}

To evaluate the uncertainty in the normalization of the transfer cross section, the dependence on a) the nucleon optical potential and b) the $^{132}$Sn  mean field generating the neutron valence orbitals in $^{133}$Sn need to be understood \footnote {Note that at this beam energy excitation of the $^{132}$Sn core is negligible.}. Concerning a), the data were reanalyzed using optical model parameters from Bechetti-Greenlees (BG) \cite{Bec69}.  FR-ADWA analysis requires nucleon potentials, whereas DWBA uses deuteron optical potentials, the latter being far more ambiguous.
The results of the FR-ADWA-BG calculations are shown by the thin dot-dashed lines in Fig.~\ref{ang} after scaling to the data. The resulting normalization factors are shown in Table~\ref{tab1}. Apart from minor changes in the shape of the distributions, differences in the normalization are 15\% for the ground state, 13\% for the state at 854 keV, 11\% for the state at 1363 keV and 10\% for the state at 2005 keV.
Note that neither BG nor CH89 have been developed for neutron-rich nuclei nor for reactions at $5$~MeV/nucleon. CH89 is a more modern interaction and in the last few years has been successfully used to describe reactions with rare isotopes, including reactions at low energies. The comparison between CH89 and BG provides an upper limit for the uncertainty introduced by the nucleon optical potentials. These could be significantly reduced by measuring proton optical potentials on neutron-rich nuclei in this mass region. 

Concerning b), FR-ADWA calculations were repeated using radius $r=1.2$~fm and diffuseness $a=0.6$~fm. The shapes of the angular distributions do not change when the geometry of the bound state is varied.  While spectroscopic factors increase by up to 40\%, the extracted ANCs remain essentially the same,
confirming that this reaction is mostly peripheral and therefore not sensitive to details of the wavefunction in the interior.

Although FR-ADWA starts from the three-body Hamiltonian $n+p+Sn$ and goes well beyond DWBA, it does not provide an exact solution to the three-body problem. In order to treat the three-body dynamics fully, a complete Faddeev solution would be necessary. Faddeev methods in momentum space (usually referred to as AGS for  Alt, Grassberger and Sandhas)  have been applied to nucleon-transfer reactions \cite{Del09} and have been used to determine uncertainties associated with FR-ADWA \cite{Nun11}.  Given the technical difficulties of AGS methods in the treatment of the Coulomb part of the interaction when heavy nuclei are involved, calculations for $^{132}$Sn are not available at present.

\section{\label{sec:discussion}Discussion\protect\\ }
The spectroscopic factors for the four states in $^{133}$Sn extracted here, using finite range ADWA, are all compatible with unity, within experimental and theoretical uncertainties, as summarized in Fig.~\ref{angdist} and Table~\ref{tab1}. This was also reported in the previous analysis although, as pointed out earlier, the DWBA analysis in \cite{Jon10} has larger uncertainties due to the ambiguities in the deuteron potential. Since the absolute values of S depend strongly on the choice of the single-particle parameters, $^{208}$Pb(d,p)$^{209}$Pb  was also analyzed \cite{Jon10} with a consistent set of  parameters. Extracted spectroscopic factors were found to also be  close to one.  This serves to further strongly validate the $N=82$ shell closure as very robust in this region and $^{132}$Sn as a good doubly-magic nucleus.

While it is accepted that spectroscopic factors are model dependent, ANCs are largely insensitive to the parameterization of the geometry of the bound state and the optical model parameters.  For this reason they are more reliable quantities for use in analyzing peripheral reactions, such as transfer reactions at energies near the Coulomb barrier. Because the $^{132}$Sn(d,p) reaction reported here is very peripheral, ANCs can be extracted with virtually no uncertainties coming from the description of the overlap function. Therefore, the experimental uncertainties, as given in Table~\ref{tab2}, reflect the total uncertainties in the ANCs.  This is the first time that ANCs for states in this region of the nuclear chart have been determined.

Spectroscopic factors are not observables  but rather are deduced from cross sections.  Therefore, it is crucial to understand the values presented in Table~\ref{tab1} within a larger context.  There has been discussion in the literature about extraction of spectroscopic factors from different types of reactions.   Spectroscopic factors extracted from (e,e'p) reactions are significantly reduced compared to those from transfer measurements
using standard DWBA, as noted by Kramer et al \cite{Kra01}. Similarly, knockout experiments lead to reduced values of S for all but the most weakly-bound nucleons \cite{Gad08}. The apparent disagreement between transfer and knockout is reduced when carefully  taking into account all sources of uncertainty in the reaction theory \cite{Nun11}. 

When the extreme quenching of spectroscopic factors originating from knocking out deeply bound particles was observed \cite{Gad08}, one might have assumed that these would correspond to short range correlations missing from the shell model. More recently, structure models have found it hard to corroborate this assumption. 
Large-scale shell model calculations with particle-vibrational couplings \cite{Bar09} were able to reproduce the large reductions and thus suggest that the large reduction of strength cannot be interpreted as coming uniquely from short-range correlations. Coupled-cluster calculations for spectroscopic factors for proton removal \cite{Jen11} also show significant quenching of spectroscopic factors when coupling to the continuum is included, again reinforcing that the reduction is caused by more complex mechanisms. 
%The effective field theory perspective \cite{fur02} is also worth considering. 

While reductions in S compared to large-scale shell model calculations are generally not observed when using a standard analysis of transfer reactions \cite{Lee07}, it is clear that transfer reactions have a strong dependence on the single-particle parameters chosen to describe the many-body overlap function \cite{Muk05}.  Lee and collaborators observed that when the geometry of the bound state potential is constrained using radii from Hartree-Fock calculations, the values of S are reduced and can be made consistent with those from (e,e'p) within error bars \cite{Lee06}. Another aspect that has been considered is the non-locality in the bound-state interaction, which is known to reduce spectroscopic factors \cite{Kra01}. Alternate forms of determining the overlap function have been proposed \cite{Tim09}. In the current work, a Woods-Saxon potential with standard radius and diffuseness was used for the bound state  and no non-locality corrections were introduced.

\section{\label{sec:conclusions}Conclusions\protect\\ }
The elastic scattering and neutron-transfer onto a 630-MeV beam of $^{132}$Sn have been studied.  The elastic scattering data were measured at angles where the Rutherford scattering dominated and nuclear effects accounted for less than $8\%$ of the reaction cross section.  This allowed for a reliable normalization of the transfer data, but the angular range was too narrow to constrain the deuteron optical potential. Four excited states in $^{133}$Sn were populated in the $^{132}$Sn(d,p) reaction in inverse kinematics, the ground, 854- and 2005-keV states that had been previously observed, as well as a newly observed state at 1363 keV.  The analysis of the angular distributions support 7/2$^-$ - 2f$_{7/2}$ and 3/2$^-$ - 3p$_{3/2}$ assignments to the ground and first excited states, respectively.  The neutron-transfer data were analyzed within the finite range adiabatic wave method. Within this approach, the deuteron wave is determined from nucleon, rather than deuteron, optical potentials and two choices of nucleon optical potentials were used. Both the spectroscopic factors and the asymptotic normalization coefficients were extracted, the latter being independent of the model used to describe the bound state in $^{133}$Sn.  The spectroscopic factors for all of these states are consistent with S~=~1 for the proposed assignments.  For a standard parameter choice for the mean field of the valence neutron in $^{133}$Sn, results are consistent with $^{132}$Sn being an excellent closed core.

\begin{acknowledgments}
This work was supported by the US Department of Energy
under contract numbers DE-FG02-96ER40955 (TTU),  DE-AC05-00OR22725 (ORNL),
DE-FG02-96ER40990 (TTU), DE-FG03-93ER40789 (Colorado School of Mines),
DE-FG02-96ER40983 (UT), DE-AC02-06CH11357
(MSU), the TORUS collaboration DE-SC0004087, the National Science Foundation under contract numbers
NSF-PHY0354870 and NSF-PHY0757678 (Rutgers) and NSF-PHY-0555893
(MSU), and the UK Science and Technology Funding Council under contract
number PP/F000715/1.  This research was sponsored in part by the National Nuclear Security Administration
under the Stewardship Science Academic Alliances program through DOE Cooperative Agreement
DE-FG52-08NA28552 (Rutgers, ORAU, MSU).
\end{acknowledgments}

\appendix*
\section{Data}
Data are tabulated below for the $^{132}$Sn(d,d)$^{132}$Sn and $^{132}$Sn(d,p)$^{133}$Sn reactions in inverse kinematics.  The elastic scattering data shown in Fig.\ref{elastics} are presented in Table \ref{tabelastics} as a ratio of the measured cross section  to that calculated for pure Rutherford scattering.   The data from the neutron-transfer reaction, as shown in Fig.\ref{angdist}, were tabulated in the supplementary information of reference \cite{Jon10} and are repeated here in Tables \ref{gs}, \ref{854}, and \ref{hiex} for completeness.

\begin {table}[ht]
\caption{\label{tabelastics}
Data from the elastic scattering of $^{132}$Sn on a deuteron target.  The cross section is expressed as a ratio to the calculated Rutherford cross section. (The uncertainties quoted here are purely statistical.  The overall uncertainties are dominated by systematics at a level of 5\%)}
\begin{ruledtabular}
\begin{tabular}{cc}
\textrm{$\theta _{CM} (deg.)$}&
\textrm {Ratio to Rutherford cross section}  \\
\colrule
28.37  & 1.042  $\pm$ 0.010\\
29.36  & 1.066  $\pm$ 0.010\\
30.36  & 1.061  $\pm$ 0.010\\
31.35  & 1.016  $\pm$ 0.011\\
32.35  & 1.005  $\pm$ 0.012\\
33.35  & 0.987  $\pm$ 0.013\\
34.34  & 1.000  $\pm$ 0.014\\
35.34  & 1.019  $\pm$ 0.014\\
36.33  & 0.998  $\pm$ 0.015\\
37.33  & 0.970  $\pm$ 0.015\\
38.33  & 1.002  $\pm$ 0.016\\
39.32  & 0.948  $\pm$ 0.018 \\
\end{tabular}
\end{ruledtabular}
\end{table}

\begin {table}
\caption{\label{gs}
Differential cross sections measured for the $^{132}$Sn(d,p)$^{133}$Sn$_{g.s.}$ reaction.
}
\begin{ruledtabular}
\begin{tabular}{cc}
\textrm{$\theta _{CM} (deg.)$}&
\textrm {d$\sigma$/d$\Omega$(mb/sr)}  \\
\colrule
28.8  & $4.3\pm0.7$\\
31.7  & $6.9\pm0.8$\\
34.7  & $6.3\pm0.7$\\
37.6  & $8.2\pm0.9$\\
42.1  & $6.9\pm1.1$\\
47.5  & $9.3\pm1.5$\\
54.5  & $8.8\pm1.5$\\
59.8  & $8.8\pm1.3$\\
64.1  & $8.4\pm0.9$\\
\end{tabular}
\end{ruledtabular}
%\end{table}

%\begin {table}[ht]
\caption{\label{854}
Differential cross sections measured for the $^{132}$Sn(d,p)$^{133}$Sn reaction to the 854-keV excited state.
}
\begin{ruledtabular}
\begin{tabular}{cc}
\textrm{$\theta _{CM} (deg.)$}&
\textrm {d$\sigma$/d$\Omega$(mb/sr)}  \\
\colrule
26.4  & $11.5\pm1.2$\\
29.1  & $11.7\pm1.1$\\
32.1  & $13.1\pm1.1$\\
43.8  & $16.5\pm1.7$\\
50.2  & $15.1\pm1.5$\\
61.6  & $11.9\pm1.3$\\
\end{tabular}
\end{ruledtabular}
%\end{table}
%\begin {table}[ht]
\caption{\label{hiex}
Integrated cross sections measured for the $^{132}$Sn(d,p)$^{133}$Sn reaction to the 1353- and 2005-keV excited states.
}
\begin{ruledtabular}
\begin{tabular}{ccc}
\textrm{E$_x$(keV)}&
\textrm{Range of $\theta _{CM} (deg.)$}&
\textrm {d$\sigma$/d$\Omega$(mb/sr)}  \\
\colrule
1363  & 33.3 to 54.0 &  $8.7\pm1.7$\\
2005 & 48.4 to 54.7  & $11.3\pm1.9$\\
\end{tabular}
\end{ruledtabular}
\end{table}

The uncertainties in the spectroscopic factors and ANCs originating from ambiguities in reaction calculations, in particular from the choice of optical potential,  are not included in Tables \ref{tab1} and \ref{tab2}.  As the relevant measured quantity for extracting spectroscopic factors is the differential, or integrated, cross section  the main sources of experimental uncertainty are: the extraction of population strength from fitting the Q-value spectra (Q-value fitting) and normalization of the data using the elastic scattering data as shown in Fig. \ref{elastics} and Table \ref{tabelastics}.  There is an additional source of uncertainty arising from the fitting of the calculated angular distribution to the data, which includes the statistical uncertainty in each data point.  A breakdown of the contributions to the uncertainties in spectroscopic factors and ANCs is shown in Table \ref{tab3}.  These contributions, as well as the total uncertainty, are given as a percentage of S or the ANC.

\begin {table*}
\caption{\label{tab3}
Sources of experimental uncertainty in spectroscopic factors and ANCs.
}
\begin{ruledtabular}
\begin{tabular}{ccccc}
&  \multicolumn{4}{c}{Percentage Uncertainty}\\
\cline{2-5}
%& \multicolumn{1}{c} {Q-value} &
 %&  \multicolumn{1}{c} {Fitting to} &\\
\textrm{E$ _x (keV)$}&
\textrm {Q-value fitting} & 
\textrm{Normalization} &
\textrm{Fitting to angular distribution} & 
\textrm{Total} \\
\colrule
0 			& 4.0		& 5 	& 5 	& 8 \\

854 			& 3.5 	 & 5   & 5 	& 8 \\

1363 	& 6.4		& 5	& 20 	& 22 \\

2005 		& 7.0 	& 5	& 20 	& 22 \\
\end{tabular}
\end{ruledtabular}
\end{table*}

\bibliography{132Sn_PRC}% Produces the bibliography via BibTeX.

\end{document}